\documentclass[12pt,A4paper]{article} 
\usepackage{indentfirst,bm,geometry,amsmath} 
\usepackage{graphicx}
\usepackage{float}
\usepackage{caption} 
\usepackage{tikz}
\usepackage{enumitem}
\usepackage{booktabs,setspace} 
\usepackage{threeparttable} 
\usepackage{multirow} 
\usepackage{diagbox} 
\usepackage{array} 
\usepackage{rotating} 
\usepackage{amssymb}
\usepackage{hyperref} 
\usepackage{xurl}    %
\usepackage{url}     %
\usepackage[linewidth=1pt]{mdframed}
\usepackage[round,authoryear]{natbib}
\usepackage{amsthm}
\usepackage{subcaption}
\newtheoremstyle{joepresult}
  {6pt}   
  {6pt}   
  {\itshape} 
  {}      
  {\normalfont} 
  {.}     
  {0.5em} 
  {}      

\theoremstyle{joepresult}
\newtheorem{result}{Result}

\hypersetup{         
    colorlinks=true, 
    linkcolor=blue,  
    filecolor=magenta, 
    urlcolor=cyan,   
    citecolor=blue  
}

\title{

Do Humans Bargain Differently with AI? \\ Evidence from Alternating-Offer Games\footnote{This research has benefited from the financial support of (a) the Joint Usage/Research Center, the Institute of Social and Economic Research (ISER), and the University of Osaka; (b) Grants-in-Aid for Scientific Research (Nos. 23H00055 and 25H00388) KAKENHI from the Japan Society for the Promotion of Science; and (c) the Support for Pioneering Research Initiated by the Next Generation program of the Japan Science and Technology Agency (No. JPMJSP2138). We gratefully acknowledge the support of Satsuki Yamada in conducting the experiment. Data will be made publicly available before acceptance.

}}

\author{Yuhao Fu\thanks{Graduate School of Economics, University of Osaka. E-mail: u889037j@ecs.osaka-u.ac.jp} \and Nobuyuki Hanaki\thanks{Corresponding author. Institute of Social and Economic Research, University of Osaka, and University of Limassol. E-mail: nobuyuki.hanaki@iser.osaka-u.ac.jp} \and
Haitao Wang\thanks{A non-academic institution. E-mail: wangtiedan2@yahoo.com}
}

\begin{document} 

\maketitle
\begin{abstract}
Artificial intelligence increasingly participates in economic interactions not only as a tool, but also as an autonomous bargaining counterpart negotiating on behalf of firms, platforms, and consumers. Yet little is known about how humans respond psychologically and strategically when bargaining with such agents in dynamic settings. We study this question in a laboratory experiment using a three-stage alternating-offer bargaining game in which participants negotiate in real time with either another human or a GPT-based AI agent. We also introduce a human-beneficiary condition in which the AI agent's earnings may affect another participant’s payment. Agreements are not reached earlier in human–human bargaining than in human–AI bargaining, but they are reached significantly earlier when the AI’s payoff affects another participant's payoff. Human proposers offer more to human opponents than to AI agents, whereas responders become significantly more willing to accept unfair AI offers when AI earnings may benefit another human. These findings suggest that fairness and reciprocity toward AI are weaker and more conditional than toward humans, but partially re-emerge when AI outcomes affect real people. The results have implications for the design of AI negotiation systems and broader human–AI economic interactions.

\end{abstract}

\hspace{0.4cm}\textbf{Keywords:} ChatGPT; bargaining game; human--AI interaction; social preferences


\medskip

\hspace{0.4cm}\textbf{JEL:} C91; C92; D83; D91 

\newpage
\doublespacing

\section{Introduction}
\label{section1}
\noindent Artificial intelligence (AI) is increasingly entering domains once dominated by human-to-human interaction. Algorithms already recommend prices, allocate resources, screen applicants, and advise consumers. More recently, large language model (LLM)-based systems have begun to act not only as decision aids, but also as autonomous bargaining counterparts that negotiate directly with humans. For example, AI negotiation systems have already begun to appear in real-world markets, including Walmart's use of AI chatbots to negotiate prices and terms with suppliers \citep{VanHoekEtAl2022Walmart}, ASOS's AI-powered bargaining bot Nibble for customer price haggling \citep{Faithfull2024Nibble}, and Xianyu's LLM-empowered bargaining agent for buyer--seller negotiations in China's online second-hand market \citep{kong2025fishbargain}. In parallel, AI research has developed human--agent negotiation platforms such as IAGO and adaptive negotiation agents that bargain with real human participants \citep{lin2010can,mell2016iago,keskin2025adaptive}.

As LLM-based AI systems continue to advance and increasingly serve as bargaining counterparts, it is important to understand how humans behave when interacting strategically with such systems, and whether that behavior differs from what is observed in traditional human--human bargaining. Although many studies have examined the bargaining ability and strategic behavior of LLM-based agents \citep{guo2023gpt,brookins2023playing,xia2024measuring,bianchi2024well,davidson2024evaluating}, less is known about how humans behave when interacting with such agents in bargaining environments, especially in dynamic settings rather than in one-shot interactions. This distinction matters because many real-world negotiations involve sequential proposals and responses rather than a single take-it-or-leave-it decision. In this sense, an alternating-offer bargaining framework, in which players make proposals in turn and costly delay affects payoffs, provides a more suitable setting for studying human--AI bargaining \citep{Rubinstein1982,OchsRoth1989}. Our first research question is therefore \textit{whether humans bargain differently with AI agents than with other humans in alternating-offer games.}

Another important consideration is that many AI bargaining agents used in commercial settings do not bargain for themselves. Instead, they negotiate on behalf of a firm, a seller, or another human stakeholder \citep{luo2006acquiring,IBM2026AgenticCommerce,Visa2026B2AI}. In such cases, the payoff allocated to the AI agent is not merely a machine payoff, but is ultimately linked to the interests of another human or organization. This feature distinguishes many real-world AI negotiations from interactions with a purely artificial counterpart whose payoff has no social consequences.

This distinction may matter because experimental studies suggest that people often respond differently to non-human counterparts than to human counterparts, and that social-preference considerations may be weaker when the counterpart is a purely mechanical agent whose payoff has no human consequence \citep{chugunova2020we,vonSchenkKlockmannKobis2025,liefooghe2023effects}. At the same time, recent evidence suggests that such considerations can partly re-emerge when the algorithm's payoff ultimately benefits a human beneficiary \citep{ozkes2024ultimatum}. Whether a similar pattern arises in dynamic bargaining environments involving LLM-based AI agents remains unclear. Our second research question is therefore \textit{whether linking the AI agent's payoff to another human affects how humans bargain with it.}


Throughout the paper, we use ``social-preference considerations'' broadly to refer to fairness concerns and concerns about the payoff consequences of one's decisions for others. In our setting, these considerations may appear differently across roles: in proposers' offers on the one hand, and in responders' acceptance or rejection of unequal allocations on the other.

More broadly, our questions can therefore be understood as asking \textit{whether, relative to human--human bargaining, some social-preference considerations become weaker in human--AI bargaining}, and \textit{whether such considerations may partly re-emerge when the AI agent's payoff is linked to another human beneficiary}. With this context in mind, we conducted a laboratory experiment in which participants bargained in real time with either another human participant or an LLM-based AI agent in an alternating-offer bargaining setting based on the design of \citet{OchsRoth1989}. We also introduced a human-beneficiary manipulation, under which the AI agent's payoff could affect another participant's final payment.

The results show that agreement timing does not differ significantly between human--human and human--AI bargaining, but within human--AI bargaining, agreements are reached significantly earlier under the human-beneficiary condition. More importantly, we find a clear asymmetry in treatment effects between proposer and responder roles, highlighting a role-dependent difference in behavior. On the proposer side, human participants make more generous offers when bargaining with another human than when bargaining with an AI agent, while the human-beneficiary manipulation does not significantly increase generosity relative to bargaining with a pure AI agent. On the responder side, by contrast, human participants do not differ significantly in their willingness to accept unfair offers when facing a human proposer rather than a pure AI proposer. However, when the AI agent's payoff is linked to another human beneficiary, responders become significantly more willing to accept unfair offers proposed by the AI agent. Thus, the human-beneficiary manipulation affects bargaining behavior mainly through responders' acceptance decisions rather than through proposers' opening offers.

We then examine several possible explanations for this asymmetry, focusing on first-mover advantage (hereafter, FMA) \citep{Rubinstein1982,OchsRoth1989}, learning, and the roles of prior and posterior beliefs. Although learning and belief-based explanations provide useful evidence, they do not appear to be the main drivers of the asymmetric treatment effects. FMA offers a more direct interpretation. We find a substantial FMA in all three treatments, and this advantage is stronger in human--AI bargaining than in human--human bargaining. This strong FMA may constrain the behavioral expression of social-preference considerations on the proposer side, making the human-beneficiary manipulation insufficient to generate a significant increase in offers. On the responder side, by contrast, the accept-or-reject decision makes the social consequences of unfair offers more directly relevant. This may explain why the human-beneficiary manipulation has a clearer effect on responders' acceptance decisions.

This study makes three main contributions. First, it extends the experimental literature on human--AI bargaining by studying a real-time finite-horizon alternating-offer game with delay costs and alternating proposer--responder roles. This allows us to study human interaction with an LLM counterpart in a dynamic bargaining environment. Second, the study introduces an experimental setting that more closely reflects emerging forms of human--AI bargaining, in which AI is not merely used as a decision aid but acts as an interactive bargaining counterpart whose payoff can be tied to the interests of another human or organization. Third, the findings deepen our understanding of how humans respond psychologically and strategically to AI counterparts in bargaining environments, and have implications for the design, deployment, and governance of AI bargaining systems. In particular, they can inform the development of AI negotiators that are more transparent, predictable, and aligned with human interests.

The remainder of this paper is organized as follows. Section 2 reviews previous studies on experimental evidence from alternating-offer bargaining, human--AI bargaining, and LLM-based strategies in bargaining contexts. Section 3 presents the experimental design and hypotheses. Section 4 reports the main results. Section 5 discusses possible mechanisms and interpretations, and Section 6 concludes.

\section{Literature Review}
\label{section2}

\noindent We review three strands of related literature. First, we summarize experimental evidence on alternating-offer bargaining, which provides the theoretical and experimental foundation for our design. Second, we review studies on human--AI bargaining and negotiation, focusing on how humans respond to algorithmic or AI counterparts. Third, we discuss recent work on LLM-based bargaining behavior, which examines the strategic capabilities and behavioral patterns of LLM-based agents themselves.

\subsection{Alternating-offer Bargaining: Experimental Evidence}

\noindent The theoretical benchmark for alternating-offer bargaining is provided by \citet{Rubinstein1982}, who analyzes an infinite-horizon bargaining game in which two players make offers in turn and delay is costly. Under complete information and standard assumptions, the model predicts immediate agreement, with the division of surplus determined by the players' relative patience. This framework has become the canonical benchmark for the analysis of dynamic bargaining in the literature.

A central experimental contribution is \citet{OchsRoth1989}, on which our design directly builds. Using finite-horizon alternating-offer bargaining games with different maximum horizons and combinations of discount factors, they show that observed behavior does not closely match the standard subgame perfect equilibrium (SPE) prediction when bargainers are assumed to care only about monetary payoffs. In particular, they document a clear FMA, as well as patterns of rejected offers and counteroffers that suggest the importance of fairness-related considerations beyond pure monetary self-interest. 

Subsequent experimental studies have extended this line of work in two main directions. One set of studies examines why bargaining behavior deviates from the exact SPE prediction. For example, \citet{weg1990two} find that although most agreements are reached immediately, their distribution is better explained by equality-based heuristics than by the equilibrium benchmark. Similarly, \citet{weg1996bargaining} show that changes in outside options move demands qualitatively in the direction predicted by subgame perfect equilibrium, but observed demands remain systematically different from exact equilibrium levels. Another set of studies examines how institutional details affect bargaining outcomes. \citet{sonnegaard1996determination} shows that proposer behavior is robust to alternative procedures for assigning the first-mover role, but responds to framing and stronger monetary incentives. More recently, \citet{heggedal2024discounting} compare three laboratory implementations of discounting in finite-horizon alternating-offer bargaining games and find no sensitivity to the number of periods. Changes in discount factors have only small and mixed effects, but disagreement occurs more frequently under effective-discounting and bargaining-delay procedures than under shrinking-pie bargaining.

In this study, we adopt the three-stage design and the discount-factor combination $(0.6,0.4)$ from \citet{OchsRoth1989}, and extend it by introducing an LLM-based AI bargaining counterpart. Human participants bargain with this AI agent in real time in a laboratory setting. This design allows us to examine whether familiar behavioral patterns in alternating-offer bargaining continue to hold when the counterpart is an AI agent rather than another human.

\subsection{Human--AI Bargaining}

\noindent Although few experimental studies have examined bargaining between humans and LLM-based agents, experimental studies of human--machine bargaining do exist, including laboratory and online negotiation experiments using alternating-offer protocols \citep{lin2010can,mell2016iago,keskin2025adaptive}. Most of this work, however, comes from the automated-negotiation and HCI literatures, and typically examines multi-issue or chat-based negotiation rather than the standard finite-horizon alternating-offer bargaining game used in experimental economics.

Closer to our setting, a growing economics literature examines how humans respond to algorithmic or AI bargaining counterparts in simpler one-shot bargaining environments. For example, \citet{erlei2022s} document substantial aversion to autonomous AI bargaining counterparts, with many responders preferring human opponents even at a monetary cost. In one-shot ultimatum bargaining, \citet{ozkes2024ultimatum} find that subjects do not strongly differentiate between human and algorithmic opponents overall, but are more willing to forgo higher payoffs when an algorithm’s earnings benefit a human beneficiary. Using a repeated ultimatum game, \citet{borthakur2025inequity} further show that participants reject disadvantageous offers from AI more often than comparable offers from humans, but are less likely to reject advantageous offers from AI. Rather than indicating a uniform shift in behavior, these studies suggest that social preferences such as fairness in human–AI bargaining are often weaker, asymmetric, or more context-dependent than in human–human bargaining. This perspective is consistent with experimental evidence showing that fairness concerns and conflict behavior evolve dynamically and depend on expectations and interaction processes \citep{Hyndman2023,XueSitziaTurocy2023}.

Other related studies examine AI bargaining in more applied negotiation settings. \citet{ShenJin2024BargainingAlgorithms} show in scenario-based consumer negotiation experiments that people make smaller adjustments to their counteroffers when bargaining with algorithms than with humans because they perceive algorithmic offers as more accurate and better informed with the effect especially pronounced among participants with lower socioeconomic background. In a related vignette study on employment negotiations, \citet{sondern2025employment} find that participants expect lower trust and less positive subjective value when negotiating with an AI counterpart rather than a human counterpart, and that presenting the AI with an avatar does not eliminate this difference. In addition, \citet{chen2026haggling} study supply-chain negotiations in which human retailers bargain with LLM suppliers over wholesale price and quantity. Across most conditions, human--LLM outcomes resemble established human--human benchmarks. However, when retailers bear inventory risk and bargaining is limited to structured offer exchanges, LLM suppliers secure higher wholesale prices and shift surplus toward themselves.

Overall, the literature reviewed in this subsection suggests that both counterpart identity and the social consequences of the AI agent’s payoff may matter for bargaining behavior. What remains unclear is whether similar patterns arise when humans bargain in real time with an LLM-based counterpart in a finite-horizon alternating-offer game.

\subsection{LLM-based AI Strategies in Bargaining}

\noindent Since the launch of ChatGPT, a growing literature has examined how LLM-based agents behave in strategic games and bargaining-related environments. Rather than focusing on human responses to AI counterparts, this line of work primarily studies the strategic capabilities and behavioral patterns of the models themselves. The emerging evidence suggests that LLMs are capable of participating in bargaining, but that their behavior remains imperfect, model-dependent, and sensitive to prompt and task structure. For example, \citet{bianchi2024well} show that LLMs can sustain multi-turn negotiation, but also exhibit a range of irrational bargaining behaviors. \citet{kwon2024llms} systematically evaluate multiple dimensions of LLM negotiation ability and find that stronger models, such as GPT-4, generally perform better, though important weaknesses remain in generating contextually appropriate and strategically advantageous responses. More recently, \citet{affonso2026large} compares 25 models across a large set of canonical games and documents substantial heterogeneity in strategic and bargaining-related behavior across model families. In a related direction, \citet{sinha2026language} show that bargaining outputs can also vary with the language of prompt, with average initial offers and surplus allocation shifting across linguistic framings.

Our study differs from this work by focusing on human behavior rather than model benchmarking. At the same time, because real-time alternating-offer interaction more closely resembles the form in which AI negotiation systems are likely to be deployed in practice, our design also provides evidence on how an LLM-based bargaining agent performs when used as an actual counterpart in an experimental economics setting.

\section{Experimental Design}
\label{section3}

\subsection{Procedure}

\begin{figure}[tb]
\centering
\includegraphics[width=\linewidth]{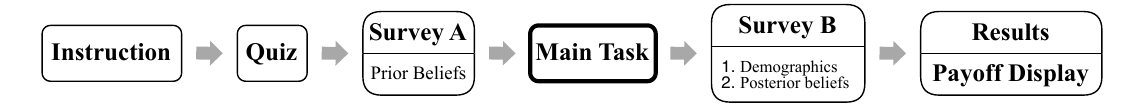}
\caption{Overall Procedure}\label{procedure}
\end{figure}

\noindent The experiment was programmed using oTree 5 \citep{chen2016otree}, and the overall procedure is shown in Figure~\ref{procedure}.

Participants first read the experimental instructions (see Online Appendix H) and were required to complete a comprehension quiz (see Online Appendix B). Only participants who correctly answered all quiz questions were allowed to proceed. After the quiz, participants completed a pre-experiment questionnaire (Survey A; see Online Appendix C.1), which elicited their prior beliefs. Participants then proceeded to the main task, where they played 10 rounds of an alternating-offer bargaining game. After completing the main task, participants filled out a post-experiment questionnaire (Survey B; see Online Appendix C.2), which collected information on demographics, including GAI experience, and posterior beliefs. Finally, a summary page displayed each participant’s payoff. All experiment screens of the main task are shown in Online Appendix I.

\subsection{Main Task}

\begin{figure}[tb]
\centering
\fbox{
\includegraphics[width=\linewidth]{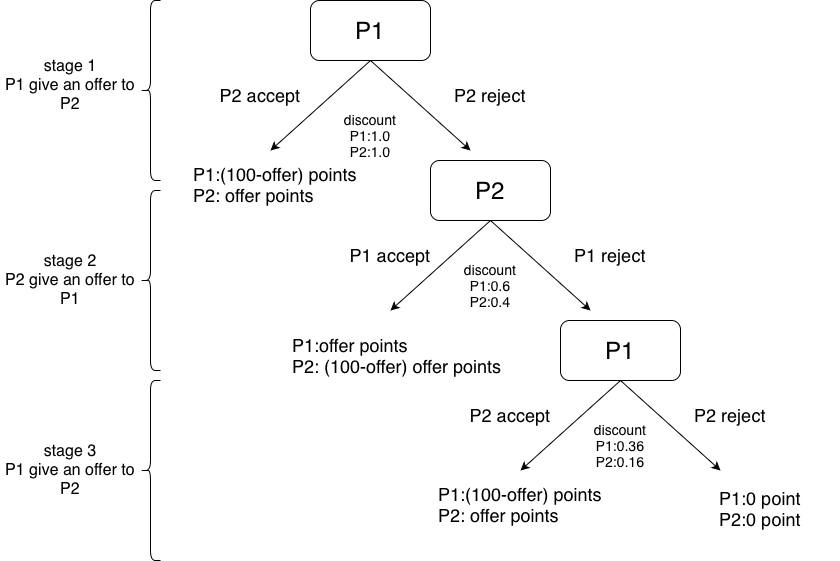}}
\caption{Main Task Structure}
\label{maintask}
\end{figure}

\noindent The main task consisted of 10 rounds of a three-stage alternating-offer bargaining game, following ``Cell 6" of the experimental design in \citet{OchsRoth1989}. The structure of the task is illustrated in Figure~\ref{maintask}.

In each round, participants were randomly matched into pairs and randomly assigned the role of either Player 1 (P1) or Player 2 (P2). They then bargained over the division of 100 points. In each round, participants were randomly re-matched and randomly reassigned to roles.

The basic rule of the alternating-offer game in each round is as follows.

In Stages 1 and 3, P1 was the proposer and P2 was the responder. In Stage 1, P1 first made an offer to P2. If P2 accepted the offer, an agreement was reached. If P2 rejected the offer, the game proceeded to Stage 2. In Stage 2, P2 was the proposer and P1 was the responder. Similarly, P2 made an offer to P1. If P1 accepted the offer, an agreement was reached. If P1 rejected the offer, the game proceeded to Stage 3, in which the roles of P1 and P2 were reversed again and became the same as in Stage 1. In Stage 3, P1 made the final offer to P2. If P2 accepted the offer, the round ended with an agreement. If P2 rejected the offer, both players received zero.

At the same time, the point payoffs received by P1 and P2 upon agreement were discounted by their respective discount factors \((\delta_1,\delta_2)\). In other words, if an offer was accepted at Stage \(t\), the agreed allocation was implemented with stage discounting: P1’s payoff was \(\delta_1^{t-1}\) times their share, and P2’s payoff was \(\delta_2^{t-1}\) times their share, where \(\delta_1 = 0.6\) and \(\delta_2 = 0.4\) in our setting. Thus, P1 was more patient than P2 and had greater bargaining power. 

\subsection{Treatments and Payment Setting}

\noindent Participants received a fixed participation fee of 500 JPY and an additional performance-based payment. In all treatments, one of the 10 rounds was randomly selected for payment. The \textit{Point Payoff} obtained in the selected round was converted into Japanese yen at a rate of 40 JPY per point, as follows,

$$\pi = 40 \times \textit{Point Payoff}.$$

\noindent We implemented three treatments that varied the identity of the bargaining counterpart and the payment rule, as follows.

\textit{T1 (Human--Human).} Participants were paired with another human participant. Payoffs were determined by the outcome of the selected round, and the \textit{Point Payoff} corresponded to the participant's own earnings in that round.

\textit{T2 (Human--AI).} Participants were paired with an AI agent (GPT model). Payoffs were determined by the outcome of the selected round, and the \textit{Point Payoff} corresponded to the participant's own earnings in that round. Participants were informed that the points allocated to the AI agent would not be used in the calculation of any participant's final payment.

\textit{T3 (Human--AI with Human Beneficiary).} Participants were paired with an AI agent. In the selected round, with 50\% probability, the \textit{Point Payoff} was equal to the participant's own earnings, and with the remaining 50\% probability, it was equal to the earnings of a randomly selected AI agent in the same role (P1 or P2) when that AI agent was matched with another participant. To minimize potential demand effects, participants were not explicitly informed that points allocated to the AI agent could affect another participant's final additional payment. Instead, they were only informed that their own final payment could, with some probability, depend on the AI agent's earnings.

\medskip
\medskip
\textit{AI Bargaining Agent}. In Treatments T2 and T3, the AI bargaining counterpart was implemented using GPT-5.4 via the Application Programming Interface (API) and embedded directly into the oTree program. Depending on the treatment condition and round-specific matching, the AI agent was assigned the role of either P1 or P2. The system prompt (see Online Appendix G.1) provided the basic rules of the alternating-offer bargaining game, while the user prompt (see Online Appendix G.2) specified the AI agent’s role, the current stage, and the history of previous stages within the same round. Accordingly, the AI agent retained memory within a round, but not across rounds. The temperature parameter was set to 1, the default setting.  No explicit reasoning parameter was specified, so the model operated under its default configuration and was used without any additional reasoning-effort setting.
Participants were informed of the AI model used, but not of the detailed prompts.

\subsection{Materials and Summary}
\noindent The design of the experiment was approved by the Institutional Review Board (IRB) of the Institute of Social and Economic Research (ISER) at the University of Osaka (\#20251003) in October 2025. The experiment was preregistered at AsPredicted (\# 277540).

\begin{figure}[tb]
\centering
\fbox{\includegraphics[width=\linewidth]{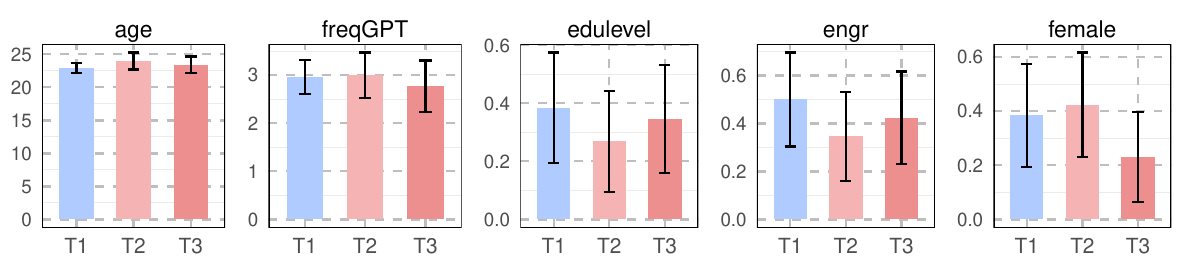}} 
\caption{Demographic Comparisons}
\vspace{-0.8em}\label{figDemo}
\caption*{\small Note: Bars report the mean of participants' reported values. Error bars denote 95\% confidence intervals across participants.}
\end{figure}

The main experiment was conducted in the laboratory of the ISER at the University of Osaka on March 10 and 17, 2026. A pilot experiment was conducted on November 19 and 20, 2025. We recruited 78 participants from the ORSEE \citep{Greiner2015ORSEE} subject pool of ISER at the University of Osaka. Each treatment included 26 participants. The sample size was determined before the main experiment based on the feasibility confirmed in the pilot experiment, and the sample sizes used in related laboratory experiments on alternating-offer bargaining. Overall, 35\% of the participants were female, and 67\% were undergraduate students. Participants were drawn predominantly from engineering (42\%), medicine (14\%), pharmacy (10\%), human sciences (8\%), and economics (4\%). In addition, 86\% of the participants reported that they regularly use ChatGPT, and only four participants reported that they had never used ChatGPT before. Variable definitions are presented in Table~\ref{demodefi}, and treatment comparisons for demographic characteristics are reported in Figure~\ref{figDemo}.

\begin{table}[t] 
\centering
\begin{threeparttable}
\caption{Demographic Statistics}
\label{demodefi}
\small{
\renewcommand{\arraystretch}{1.5} 
\begin{tabular}{lm{8cm}ccccc} 
\toprule
Var. & Definition & Min. & Max. & Avg. & S.D.\\
\midrule
age & Participant's age. & 19 & 33 & 23.4 & 2.93\\
freqGPT & Frequency of ChatGPT use; coded from 0 = ``never"  to 4 = ``multiple times per day". & 0 & 4 & 2.91 & 1.19\\
edulevel & Education level; = 1 if graduate student, = 0 if undergraduate student. & 0 & 1 & 0.33 & 0.47\\
engr & Major in engineering; = 1 if yes. & 0 & 1 & 0.42 & 0.50\\
female & Gender; = 1 if female. & 0 & 1 & 0.35 & 0.48\\
\bottomrule
\end{tabular}}
\end{threeparttable}
\end{table}

During the experiment, participants were prohibited from using their own electronic devices, including smartphones and tablets. Although all tasks were completed on laboratory computers, internet access within the experimental software was disabled.

Each session lasted about 60 minutes on average, including payment, and participants earned an average total payoff of 2324 JPY (2242 in T1, 2265 in T2, 2466 in T3).

\subsection{Hypotheses}

\noindent We first compare agreement timing between T1 (Human--Human) and T2 (Human--AI). Prior research suggests that interactions with algorithms and machine partners often elicit weaker cooperation and social preferences than interactions with human counterparts \citep{liefooghe2023effects,vonSchenkKlockmannKobis2025}. If social preferences are more salient in human--human bargaining than in human--AI bargaining, participants may be more willing to compromise early and thus reach agreement sooner when facing a human opponent. This reasoning leads to the following hypothesis:

\begin{enumerate}
[label=H1:,leftmargin=2.25em]
    \item \textit{Agreements are reached earlier with human opponents than with AI agents.}
\end{enumerate}

Because social preferences are likely to be weaker toward non-human counterparts than toward human counterparts, participants may behave less prosocially and more self-servingly in strategic settings, especially in bargaining games. In ultimatum bargaining, responders' rejection of unfair offers is commonly interpreted as a form of fairness-based punishment or negative reciprocity, since responders are willing to give up positive monetary payoffs in order to reject an unequal allocation \citep{guth1982experimental,FehrSchmidt1999,camerer1995anomalies,OosterbeekETAL2004EE}. Recent evidence from one-shot ultimatum game settings suggests that human proposers offer less when the responder's decision is made by an LLM \citep{dvorak2025adverse}, and that human responders are more likely to reject disadvantageous offers from an AI counterpart than from a human counterpart \citep{borthakur2025inequity}. Although these studies do not examine real-time, multi-stage bargaining with an embedded LLM-based agent in the laboratory, their findings suggest that similar behavioral patterns may arise in an alternating-offer bargaining game. Accordingly, we define \textit{unfair offers} as offers below 50 points to the responder and then propose the following hypotheses:

\begin{enumerate}[label=H2a:, leftmargin=*, align=left]
    \item \textit{Human proposers offer smaller shares to AI agents than to human opponents.}
\end{enumerate}
\vspace{-1.5em}
\begin{enumerate}[label=H2b:, leftmargin=*, align=left]
    \item \textit{Human responders are less willing to accept unfair offers from AI agents than from human opponents}.
\end{enumerate}

We then consider the relationship between T2 and T3. Linking the AI agent's earnings to another human participant may reintroduce social-preference considerations into human--AI bargaining. In other words, some of the social preferences that are weakened in T2, relative to T1, may partially recover once participants realize that their allocation to the AI agent may affect the payoff of another participant, even though their direct bargaining counterpart is still an AI agent. This intuition is consistent with recent evidence showing that social preferences toward AI agents become stronger when the AI's earnings benefit a human \citep{vonSchenkKlockmannKobis2025,ozkes2024ultimatum}. 

For proposers, this reasoning implies that they may offer larger shares to AI agents when the AI's earnings are linked to another human participant's payment. For responders, however, the human-beneficiary manipulation changes the social meaning of rejection. Rejecting an unfair offer may still punish the AI proposer, but it may also reduce the payoff of the human beneficiary. Thus, responders may become more willing to accept unfair offers not because fairness concerns disappear, but because rejection now has a social cost for another human. We therefore propose the following hypotheses for the human-beneficiary manipulation:

\begin{enumerate}[label=H3a:, leftmargin=*, align=left]
    \item \textit{Human proposers offer larger shares to AI agents when the AI's earnings are linked to another human participant's payment.}
\end{enumerate}
\vspace{-1.5em}
\begin{enumerate}[label=H3b:, leftmargin=*, align=left]
    \item \textit{Human responders are more willing to accept unfair offers from AI agents when the AI's earnings are linked to another human participant's payment.}
\end{enumerate}

\section{Results}
\label{section4}

\noindent This section presents the main findings of the experiment. We first examine agreement timing and the corresponding treatment effects. We then turn to opening offers, the corresponding responses, and counteroffers. Because the vast majority of agreements were reached in the first two stages, the number of observations from Stage 3 and from rounds ending without agreement is too small for separate analysis. Accordingly, we focus on behavior in the first two stages.

\subsection{Agreement Timing}

\noindent In T1, the 26 participants were randomly matched into pairs in each of the 10 rounds, resulting in 130 bargaining observations. In T2 and T3, each participant bargained with an AI agent in each round, resulting in 260 bargaining observations in each treatment. Table~\ref{tab:maxstage} reports the percentage distribution of agreement timing across stages. Bargaining cases in which no agreement was reached by the end of Stage 3 are coded as Stage 4.

\begin{figure}[tb]
\centering
\fbox{\includegraphics[width=\linewidth]{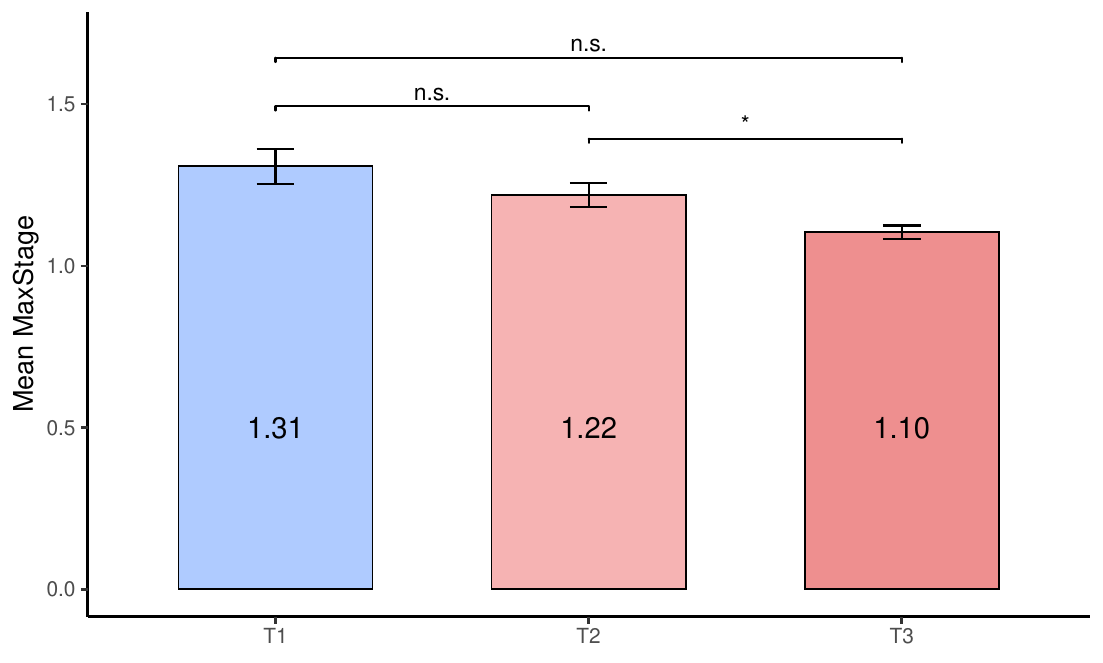}}
\caption{Mean Stage Reached Across Treatments}
\label{fig:maxstage}
\caption*{\small Note: * $p<.05$, ** $p<.01$, *** $p<.001$. ``n.s.'' means that the difference is not statistically significant at the 5\% level. Error bars denote 95\%
confidence intervals across participants. $\text{MaxStage}$ is compared across treatments using the Mann–Whitney U test. }
\end{figure}

\begin{table}[H]
    \centering
    \caption{Distribution of Agreement Timing (\%)}
    \label{tab:maxstage}
    \begin{tabular}{lccc|c}
        \hline
        Treatment & Stage 1 & Stage 2 & Stage 3 & Stage 4\\
        \hline
        T1 & 88.46 & 0.77 & 2.31 & 8.46  \\
        T2 & 85.38 & 9.23 & 3.46 & 1.92  \\
        T3 & 90.77 & 8.08 & 1.15 & 0.00 \\
        \hline
        Total & 88.15 & 7.08 & 2.31 & 2.46 \\
        \hline
    \end{tabular}
\end{table}
More specifically, most agreements were reached at Stage 1 in all three treatments. In T1, the numbers of bargaining cases ending at stages 1, 2, 3, and 4 were 115, 1, 3, and 11, respectively. In T2, the corresponding numbers were 222, 24, 9, and 5, while in T3 they were 236, 21, 3, and 0. T2 and T3 exhibit the expected monotonic decline across stages, whereas T1 shows a less regular pattern, with unusually few cases ending at Stage 2 and a comparatively larger number of bargaining failures. One possible explanation is that, in T1, bargaining cases that did not end at Stage 1 were simply harder to settle, and therefore were also less likely to reach agreement in later stages. Figure~\ref{fig:maxstage} compares the mean stage reached (MaxStage) across treatments.

There is no significant difference between T1 and T2 ($p =.528$), and thus H1 is not supported. However, agreements are reached significantly earlier in T3 than in T2 ($p =.044$), suggesting that participants may show greater consideration for the payoff consequences of the AI agent's allocation when the AI agent's payoff is linked to another human participant's payment.

\begin{result}
\label{result1}
\textit{Agreements are not reached earlier with human opponents than with AI agents.}
\end{result}

\begin{result}
\label{result2}
\textit{When bargaining with AI agents, agreements are reached earlier when the AI agent's payoff is linked to another human participant's payment.}
\end{result}

\subsection{Opening Offers}

\subsubsection{Human Proposer}
\label{sectionOpeningOffers}
\begin{figure}[tb]
\centering
\fbox{\includegraphics[width=\linewidth]{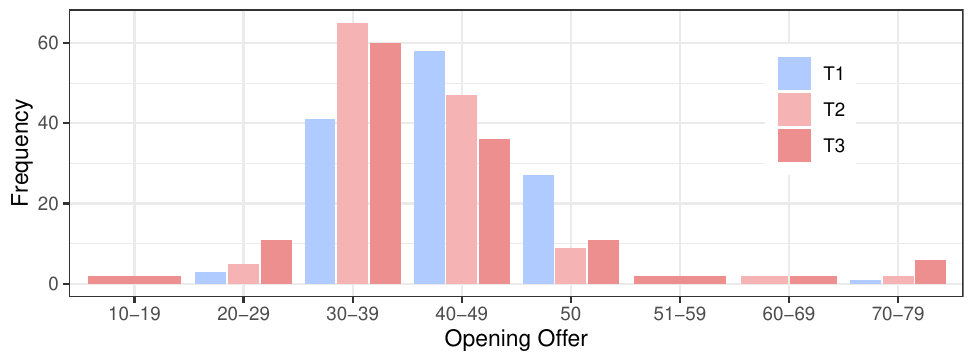}}
\caption{Human Opening Offers}
\label{fig:OpeningOffer_HM}
\end{figure}

\noindent In all treatments, half of the human participants were assigned the role of P1 in each round. Therefore, the total sample of opening offers was 130 in each treatment. The distributions of human opening offers are presented in Figure~\ref{fig:OpeningOffer_HM}. In T1, human proposers made opening offers to human opponents, whereas in T2 and T3, human proposers made opening offers to AI agents.

The figure shows that, in T1, most opening offers fall in the 40--49 range, whereas in T2 and T3, most opening offers made to AI agents fall in the 30--39 range. This pattern suggests that participants were more generous toward human opponents than toward AI agents. In particular, the 50--50 proposal, which can be interpreted as a \textit{fair allocation}, was observed 27 times in T1, compared with only 9 times in T2 and 11 times in T3. This difference further suggests that fairness considerations were stronger when participants faced human opponents than when they faced AI agents. More broadly, this pattern is consistent with the conjecture that social preferences are stronger in human--human bargaining than in human--AI bargaining.

It is also worth noting that almost no participants chose the SPE benchmark. In T3, only one participant made the SPE opening offer of 16, and only one participant proposed less than 16. All remaining opening offers were above the SPE level.

\begin{figure}[tb]
\centering
\fbox{\includegraphics[width=\linewidth]{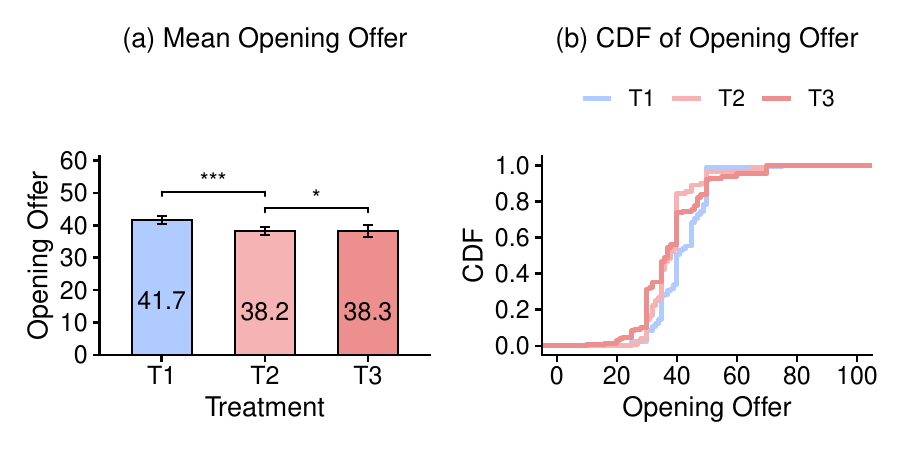}}
\caption{Comparisons of Human Opening Offers across Treatments}
\label{comparision_OpeningOffer_HM}
\caption*{\small Note: Panel (a) shows mean opening offers with 95\% confidence intervals. Panel (b) shows the cumulative distribution functions (CDF) of opening offers. * $p<.05$, ** $p<.01$, *** $p<.001$. Opening offers are compared across treatments using the Kolmogorov-Smirnov test.}
\end{figure}

Figure~\ref{comparision_OpeningOffer_HM} shows comparisons of human opening offers across treatments. Participants made higher opening offers to human opponents than to AI agents (T1 vs. T2, $p<.001$). The distribution of opening offers also differs between T2 and T3 in the Kolmogorov--Smirnov test ($p=.048$), suggesting some distributional shift under the human-beneficiary manipulation. However, as shown below, this shift does not translate into a robust increase in average opening offers once controls are included.

OLS regressions (see Table A.1 in Online Appendix A), controlling for round number and personal characteristics, show that participants in T1 made significantly higher opening offers than those in T2 across specifications. This result is consistent with H2a, suggesting that human proposers offered smaller shares to AI agents than to human opponents. By contrast, the coefficient on T3 is positive but statistically insignificant in all specifications, indicating that the human-beneficiary manipulation did not significantly increase the average opening offer relative to T2. Thus, H3a is not supported. Because most agreements were reached at Stage 1 and the number of human offers observed in Stages 2 and 3 is very limited, we summarize the findings as follows:

\begin{result}
\label{result3}
\textit{Human proposers offer more to human opponents than to AI agents, consistent with stronger social preferences in human--human bargaining than in human--AI bargaining.}
\end{result}

\begin{result}
\label{result4}
\textit{There is no significant evidence that the human-beneficiary manipulation increases offers by human proposers.}
\end{result}

\subsubsection{AI Proposer}

\noindent In treatments T2 and T3, AI agents made 130 opening offers in each treatment. The opening offers were overwhelmingly concentrated at 40 points. In T2, 128 of the 130 opening offers were 40, with only one offer of 16 and one offer of 64. Similarly, in T3, the AI offered 40 in 125 cases, while the remaining offers consisted of four cases of 64 and one case of 36.

This extremely limited variation suggests that the AI proposer’s opening behavior was highly stable. To further assess its robustness, we conducted an additional API-based exercise using the same prompt, with one added instruction asking the model to independently generate an opening offer as P1 and to explain the reasoning behind that choice. The corresponding results are presented in Online Appendix D.1.

\subsection{Responses to Opening Offers}
\subsubsection{Human Responder}

\noindent We focus on cases in which the responder to the opening offer was a human participant. This includes T1, where human responders (P2) responded to opening offers made by another human (P1), and T2 and T3, where human responders (P2) responded to opening offers made by AI agents (P1). In each treatment, this yields 130 observations. The acceptance rate of human responders to opening offers was 88.5\% in T1, 86.2\% in T2, and 97.7\% in T3. 

When the sample is restricted to unfair offers, defined as offers below 50 points to the responder, the number of observations decreases to 102 in T1, 129 in T2, and 126 in T3. For unfair opening offers, the acceptance rates of human responders were very similar in T1 (85.3\%) and T2 (86.0\%). A Fisher's exact test confirms that this difference is not statistically significant ($p = 1.000$), providing no support for H2b. By contrast, the acceptance rate in T3 rises to 97.6\%. A Fisher's exact test comparing T2 and T3 shows that this difference is statistically significant ($p < .001$), supporting H3b. 

Probit estimates (see Table A.2 in Online Appendix A), controlling for opening offers and personal characteristics, also support the findings above. In particular, the result in Specification (4), which uses the subsample of unfair opening offers, shows that the coefficient on T1 is small and statistically insignificant, whereas the coefficient on T3 is positive and statistically significant. This indicates that, relative to T2, human responders in T1 are not significantly less or more likely to accept unfair opening offers, while human responders in T3 are significantly more likely to accept them. Therefore, we have the following results.

\begin{result}
\label{result5}
\textit{Human responders do not differ significantly in their willingness to accept unfair opening offers from human and AI proposers.}
\end{result}

\begin{result}
\label{result6}
\textit{Human responders are more willing to accept unfair opening offers from AI proposers when the AI agent's earnings are linked to another human participant's payment.}
\end{result}

\subsubsection{AI Responder}

\noindent In treatments T2 and T3, AI agents responded to 130 human opening offers in each treatment. The overall acceptance rate was 84.2\%, with acceptance rates of 84.6\% in T2 and 83.8\% in T3. When the sample is restricted to unfair offers, the number of observations decreases to 117 in T2 and 109 in T3. In this restricted sample, the acceptance rates were 82.9\% in T2 and 80.7\% in T3. A Fisher's exact test confirms that this difference is not statistically significant ($p = .731$).

To further assess the robustness of these patterns, we conducted an additional API-based exercise using the same prompt. For each possible human opening offer from 0 to 100, we independently elicited the AI agent's response 100 times, together with its explanation and, when the offer was rejected, the corresponding counteroffer. The results are reported in Online Appendix D.2.

\subsection{Counteroffers}

\subsubsection{Human Counteroffers}

\noindent When a human responder (P2) rejects the opening offer proposed by P1, the responder makes a counteroffer in Stage 2. This counteroffer provides additional information on how responders trade off monetary payoff against fairness concerns \citep{OchsRoth1989}. In particular, if the responder's discounted payoff implied by the Stage-2 counteroffer is lower than the payoff that would have been obtained by accepting the opening offer, then monetary payoff alone cannot fully account for the rejection decision. In such cases, the responder appears willing to incur a material cost in order to reject an unfair offer, which is consistent with fairness-based punishment \citep{BranasGarzaEtAl2014}. Thus, materially costly rejection may reflect fairness considerations rather than purely monetary optimization.

To capture this idea, we define the \textit{FairnessGap} as
\[
\textit{FairnessGap} = \textit{Opening Offer} - 0.4 \times (100 - \textit{CounterOffer}),
\]
where \(100 - \textit{CounterOffer}\) is the responder's own payoff implied by the counteroffer, and 0.4 is the responder's discount factor in Stage 2. A positive fairness gap indicates that rejecting the opening offer is materially costly relative to the continuation payoff implied by the responder's own counteroffer.

As a supplementary exploratory analysis, we examine the subset of rejected opening offers. The number of such observations is limited, with 15 in T1, 18 in T2, and only 3 in T3. Mean counteroffers are lower in T2 than in T1 (19.39 vs.\ 27), but a Mann--Whitney U test indicates that this difference is not statistically significant ($p = .154$). Likewise, the difference in the fairness-gap measure is not statistically significant ($p = .365$).

By contrast, all rejected opening offers in T2 involve materially costly rejection, whereas this is true for 11 out of 15 rejected opening offers in T1. A Fisher's exact test indicates that this difference is statistically significant ($p = .033$). However, given the very small subsample size, especially in T3, these results should be interpreted with caution.

\subsubsection{AI Counteroffers}

\noindent Compared with human counter-proposers, AI counter-proposers appear to be more generous. We therefore examine AI counteroffers following rejected opening offers. The number of such observations is 20 in T2 and 21 in T3. Mean counteroffers are very similar across the two treatments (58 in T2 and 59.05 in T3), and a Mann--Whitney U test does not indicate a significant difference ($p = .544$).

Likewise, materially costly rejection is observed in almost all cases in both treatments: all 20 cases in T2 and 20 out of 21 cases in T3. A Fisher's exact test confirms that this difference is not statistically significant ($p = 1.000$). Additional API-based exercises using the same prompt to elicit the AI's counteroffers are reported in Online Appendix D.2.

%

\section{Discussion}
\label{section5}

\noindent Our main analysis compares T1 with T2, and T2 with T3, in order to examine how human bargaining behavior changes when the counterpart is an AI agent rather than a human, and when the AI agent's earnings are linked to another human participant's payment. The results reveal a central asymmetry between proposer behavior and responder behavior:

\begin{itemize}
    \item Human proposers are more generous toward human opponents than toward AI agents, but the human-beneficiary manipulation does not significantly increase their average generosity toward AI agents.
    \item Human responders do not differ significantly in their willingness to accept unfair opening offers from AI rather than human proposers, but they become substantially more willing to accept such offers when the AI agent's earnings are linked to another human participant's payment.
    
\end{itemize}

These asymmetric findings make it difficult to characterize the role of social preferences in human--AI bargaining using a single, uniform framework. Rather than suggesting that social-preference considerations simply weaken when the counterpart is an AI agent and are then restored once the AI's earnings affect another human, the results indicate that \textit{the influence of social preferences differs between proposers and responders}.

We therefore shift our focus in this section to several additional perspectives that may help interpret the asymmetry: (1) FMA, (2) learning, and (3) participants' prior and posterior beliefs.

\subsection{First-Mover Advantage}

\noindent In sequential bargaining, the player who makes the first proposal is generally thought to enjoy an FMA. In the standard alternating-offer framework, making the first offer allows a player to anchor the bargaining process and to exploit the fact that delay is costly for both parties \citep{Rubinstein1982,OchsRoth1989}. In the setting of this study, this advantage is further strengthened by the asymmetry in discount factors: P1 faces a lower cost of delay than P2, because \(\delta_1 = 0.6\) while \(\delta_2 = 0.4\). As a result, P1 is in a relatively stronger bargaining position than P2.

\subsubsection{Existence and Strength of FMA}

\noindent Although the opening offers discussed in Section~\ref{sectionOpeningOffers} already reflect P1's initial claim over the surplus, we further assess FMA by comparing P1's realized payoff and P1's share of the total realized payoff across treatments.

\begin{figure}[tb]
\centering
\fbox{%
\begin{minipage}{0.775\linewidth}
\centering
\includegraphics[width=\linewidth]{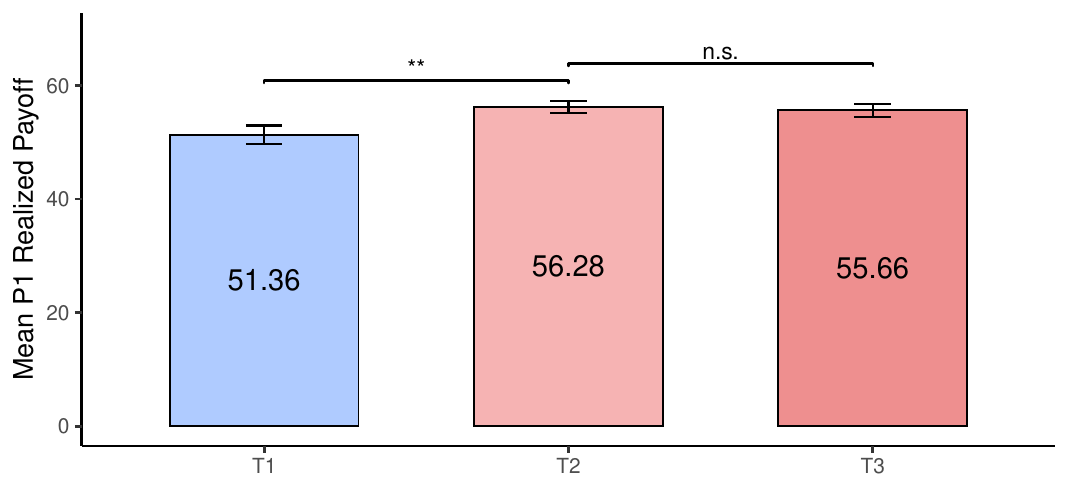}

(a) P1's Realized Payoff

\vspace{0.8em}

\includegraphics[width=\linewidth]{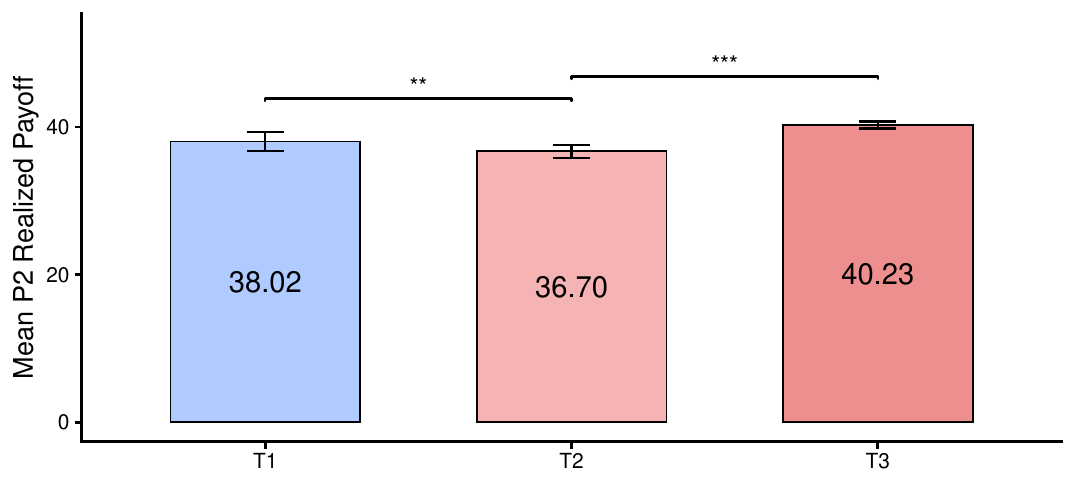}

(b) P2's Realized Payoff
\end{minipage}%
}
\caption{Comparisons of Realized Payoff across Treatments}
\label{RealizedPayoff}
\caption*{\small Note: * $p<.05$, ** $p<.01$, *** $p<.001$. ``n.s.'' means that the difference is not statistically significant at the 5\% level. Realized payoffs are compared across treatments using the Mann--Whitney U test. Error bars denote 95\% confidence intervals. In T2 and T3, only the outcomes of human P1 and human P2 are included.}
\end{figure}

\begin{figure}[tb]
\centering
\fbox{\includegraphics[width=0.9\linewidth]{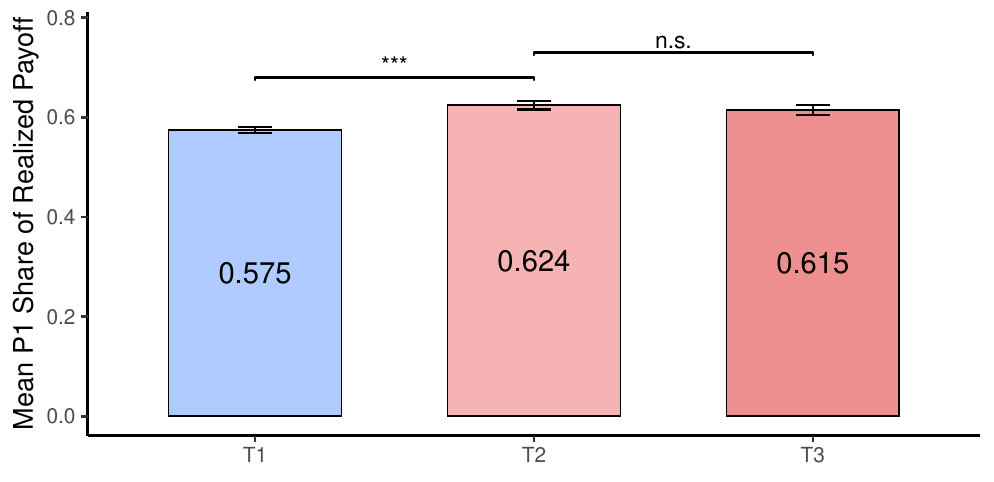}}
\caption{Comparisons of P1's Share of the Total Realized Payoff across Treatments}
\label{shareRealizedPayoff}
\caption*{\small Note: * $p<.05$, ** $p<.01$, *** $p<.001$. ``n.s.'' means that the difference is not statistically significant at the 5\% level. Error bars denote 95\% confidence intervals. Shares of the total realized payoff are compared across treatments using the Mann--Whitney U test.}
\end{figure}

Figure~\ref{RealizedPayoff} reports the realized payoffs of human P1 and human P2, where realized payoff refers to the point payoff that a human participant actually obtained in each round after discounting. This measure captures the absolute extent of FMA. Figure~\ref{shareRealizedPayoff} reports P1's share of the total realized payoff, calculated as
$$
\frac{\text{P1's realized payoff}}{\text{P1's realized payoff}+\text{P2's realized payoff}},
$$
which captures the relative extent of FMA.

First, within each treatment, P1's realized payoff is clearly and significantly higher than P2's realized payoff (Mann--Whitney U test, $p<.001$ in all three treatments). This provides direct evidence that an FMA exists in our bargaining environment.

Second, comparing across treatments, the FMA appears strongest in T2. Relative to T1, T2 yields a significantly higher realized payoff for P1 (Mann--Whitney U test, $p=.004$), a significantly lower realized payoff for P2 (Mann--Whitney U test, $p=.003$), and a significantly larger share of the total realized payoff captured by P1 (Mann--Whitney U test, $p<.001$). These results suggest that when bargaining against a pure AI agent, P1 is better able to convert the structural FMA into a larger share of the realized surplus.

Third, the comparison between T2 and T3 suggests that this pattern is partly mitigated when the AI agent's payoff is linked to another human participant's payment. Although P1's realized payoff does not differ significantly between T2 and T3 (Mann--Whitney U test, $p=.967$), its mean value declines slightly. At the same time, P2's realized payoff becomes significantly higher in T3 (Mann--Whitney U test, $p<.001$), and P1's share of the total realized payoff also declines in magnitude, although the difference is not statistically significant (Mann--Whitney U test, $p=.476$). 

We also estimate regressions of realized payoff and the share of the total realized payoff, controlling for personal characteristics and round number. The results, reported in Tables A.3--A.5 in Online Appendix A, are consistent with the graphical evidence. In particular, relative to T2, T1 is associated with a significantly lower realized payoff for P1 and a significantly lower share of the total realized payoff captured by P1, while the coefficient on T3 in the P2 realized-payoff regressions is positive but statistically insignificant.

To summarize, the pattern of FMA across the three treatments mirrors the main asymmetry in our results. The comparison between T1 and T2 shows that bargaining with a pure AI agent rather than a human is associated with a significantly larger FMA. By contrast, the human-beneficiary manipulation does not significantly reduce FMA in T3, either in terms of absolute realized payoffs or in terms of P1's share of total realized payoffs. At the same time, however, it significantly improves responders' realized payoffs, suggesting that it partly mitigates their disadvantaged bargaining position.

This pattern is also consistent with the interpretation that linking the AI agent's payoff to another participant's payment partly restores the social dimension of the interaction. When the counterpart is purely artificial, extracting a larger share may be perceived as having little social cost. When the AI's payoff potentially affects another human, this perception may be weakened, which could contribute to the observed increase in responders' willingness to accept unequal offers.

\medskip
\medskip

\subsubsection{FMA as a Possible Explanation for the Asymmetry}

\noindent The strong FMA documented above may also help explain the asymmetric findings between proposer and responder behavior in the main analysis. A possible interpretation is that the behavioral relevance of social preferences depends on the strategic role occupied by the human participant. Prior evidence from ultimatum and alternating-offer bargaining experiments suggests that responder decisions are often closely tied to fairness judgments over proposed allocations, as reflected in the rejection of unfair offers. Proposer decisions, by contrast, reflect not only fairness concerns but also strategic considerations, such as expectations about acceptance thresholds and the distribution of bargaining power \citep{guth1982experimental,camerer1995anomalies,FehrSchmidt1999,OchsRoth1989}.

This role-based interpretation can first explain why the human--AI difference appears more clearly on the proposer side than on the responder side. For proposers, counterpart identity is directly relevant to the allocation decision: they decide how much surplus to give to a human opponent or to an AI agent. If social preferences are weaker toward AI agents than toward human opponents, this difference can translate directly into lower offers in the human--AI treatment. For responders, however, counterpart identity enters the decision less directly. Responders decide whether to accept or reject a given allocation, and this decision is shaped not only by the identity of the proposer but also by the material payoff from acceptance, the desire to punish unfairness, and the expected value of continuing the bargaining process. As a result, the human or AI identity of the proposer alone may not be sufficient to generate a significant difference in acceptance behavior.

In our bargaining environment, proposers enjoy a substantial FMA, and this advantage is especially strong when humans bargain with AI agents rather than with other humans. This provides a possible explanation for the contrast between Result~\ref{result3} and Result~\ref{result4}. In the human--AI treatments, proposers are already in a favorable strategic position and can claim a relatively large share of the surplus. Therefore, even if the human-beneficiary manipulation makes social-preference considerations more salient, such concerns may be too weak to overcome the strategic incentives created by the proposer’s advantageous position. In this sense, social preferences may still operate on the proposer side, but their behavioral expression is constrained by FMA and strategic considerations.

On the responder side, \textit{the absence of FMA may make the human-beneficiary manipulation more behaviorally consequential}. Unlike proposers, responders do not occupy the structurally advantageous position created by moving first. Instead, they face a proposed allocation and decide whether to accept it or reject it. This weaker strategic position may make their decisions more sensitive to changes in the social consequences of acceptance or rejection. The human-beneficiary manipulation changes precisely this consequence. In the pure human--AI treatment, rejecting an unfair AI offer mainly functions as a response to an unfair allocation proposed by a non-human counterpart. In the human-beneficiary treatment, however, rejection may also impose a monetary cost on another human participant. Thus, the same accept/reject decision involves an additional social consequence, which may shift the relevant concern from punishing unfairness to avoiding harm to another human beneficiary.

Taken together, the asymmetry has two layers. First, the human--AI difference is more visible for proposers because counterpart identity directly affects the allocation decision. Second, the human-beneficiary manipulation is more visible for responders because it changes the social consequence of rejecting an unfair offer, whereas its effect on proposers may be muted by the strong FMA in the human--AI treatments. Thus, social preferences are not uniformly weakened or restored in human--AI bargaining; rather, their behavioral expression \textit{depends on the strategic role and decision environment}.

\subsection{Learning and Adaptation}

\noindent Since most participants were likely unfamiliar with repeated alternating-offer bargaining, especially against AI agents, some degree of learning and adaptation over the 10 rounds is to be expected. As subjects gained experience, they may have updated both their understanding of the strategic structure of the game and their expectations about counterpart behavior. In this subsection, we examine whether such learning is present, and whether it helps explain the asymmetric treatment effects observed above.

\subsubsection{Evidence of Learning Across Rounds}

\begin{figure}[tb]
\centering
\fbox{%
\begin{minipage}{0.775\linewidth}
\centering
\includegraphics[width=\linewidth]{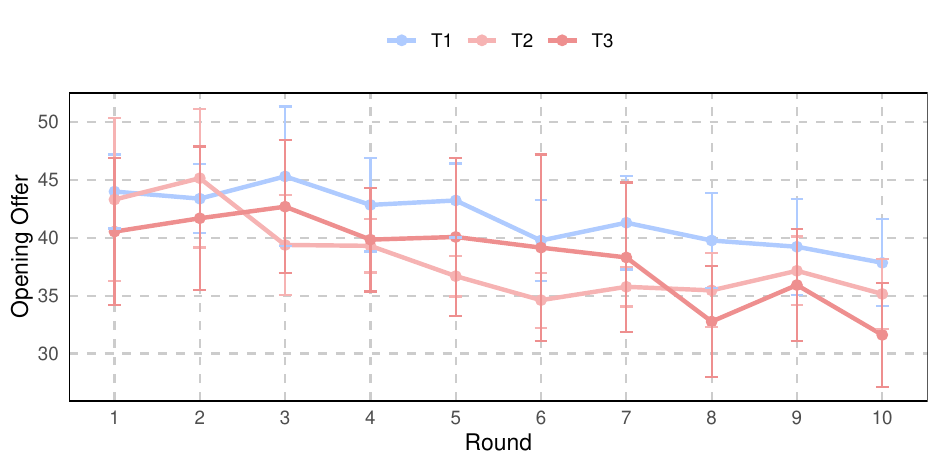}
(a) P1's Opening Offer
\vspace{0.8em}

\includegraphics[width=\linewidth]{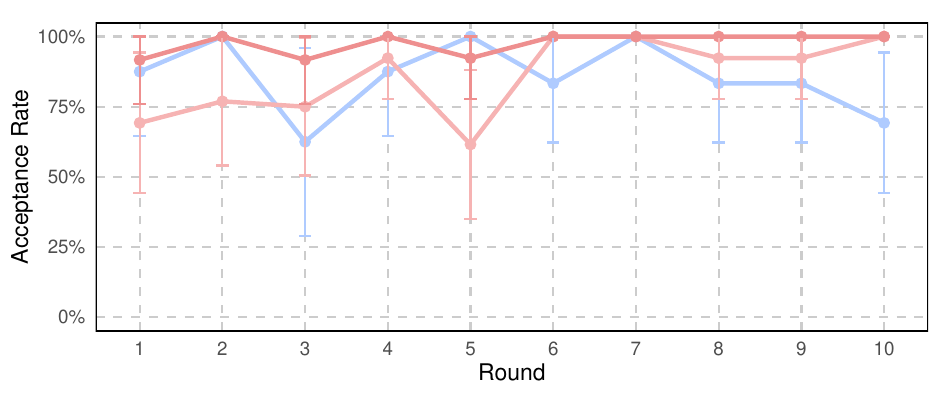}

(b) P2's Acceptance Rate Facing Unfair Opening Offers
\end{minipage}%
}
\caption{Learning and Adaptation across Rounds}
\label{fig:learning}
\caption*{\small Note: Panel (a) shows the mean opening offer of human proposers (P1) across rounds by treatment. Panel (b) shows the acceptance rate of human responders (P2) for unfair opening offers across rounds by treatment. Error bars denote 95\% confidence intervals.}
\end{figure}

\noindent Figure~\ref{fig:learning} illustrates the evolution of human proposers' opening offers and human responders' acceptance rates over the 10 rounds. Across treatments, opening offers show a clear downward trend. By contrast, on the responder side, the pattern differs across treatments. In T1, the acceptance rate for unfair opening offers changes only modestly across rounds. In T2 and T3, however, the acceptance rate fluctuates somewhat in the first half of the experiment, but remains close to 100\% in the later rounds.

The regression results reported in Online Appendix A provide further detail. In Table A.1, the coefficient on \textit{roundnum} is negative and statistically significant in the opening-offer regressions. This suggests that, over time, proposers lowered their opening offers. One possible interpretation is that participants gradually came to better understand their FMA and therefore became less generous as they gained experience. At the same time, the interaction terms between treatment indicators and \textit{roundnum} are not statistically significant, suggesting that this proposer-side learning effect does not differ substantially across treatments.

On the responder side, Table A.2 shows that the coefficient on \textit{roundnum} is positive and statistically significant in the regressions for accepting unfair opening offers. This suggests that, as rounds progressed, responders became more willing to accept unfair opening offers. However, when interactions between treatment indicators and \textit{roundnum} are introduced (Table A.6), the results suggest that this responder-side learning effect is weaker in T1 than in T2 and T3, while there is no clear difference between T2 and T3. This pattern may indicate that adaptation was stronger when participants repeatedly interacted with AI agents than when they interacted with humans.

By contrast, Tables A.4 and A.5 show that \textit{roundnum} has no significant effect on realized payoffs. This suggests that learning primarily affected bargaining behavior itself—such as offer levels and acceptance decisions—rather than improving final material outcomes. In other words, participants appear to have adapted their strategies over time, but this adaptation did not necessarily translate into higher realized earnings.

These findings indicate that learning and adaptation were present in the experiment, but they do not alter the main treatment patterns. Instead, learning appears to operate alongside the treatment effects: proposers generally became less generous over time, responders generally became more willing to accept offers, and the responder-side adaptation was somewhat stronger in AI bargaining than in human--human bargaining.

\subsubsection{Learning and the Asymmetry Between Proposers and Responders}

\noindent As shown above, behavior exhibits learning on both the proposer side and the responder side over the 10 rounds. If subjects' experience shapes how treatment differences emerge, then the strength of the treatment effects may differ between earlier and later rounds. To examine this possibility, we re-estimate the regressions for proposers' opening offers and responders' acceptance of unfair offers separately for Rounds 1--5 and Rounds 6--10. The results are reported in Table A.7 of Online Appendix A.

On the proposer side, the results in Table A.7 are broadly consistent with those in Table A.1 based on the pooled sample. In particular, the coefficient on T3 remains insignificant in both subsamples, suggesting that linking the AI agent's payoff to a human beneficiary does not significantly affect proposers' opening offers, even after accounting for potential heterogeneity by experience.

On the responder side, the pattern is more nuanced. In the first five rounds, the significance of the coefficients on T1 and T3 is broadly similar to that in column (4) of Table A.2 based on the full sample of unfair offers. In the last five rounds, however, the coefficient on T1 becomes significantly negative, indicating that responders in T1 are significantly less likely to accept unfair offers than those in T2, whereas responders in T3 remain significantly more likely to accept them. This late-round pattern further contradicts H2b, which predicted that human responders would be less willing to accept unfair offers from AI agents than from human opponents.

One possible interpretation is that learning affects how responders interpret the source of unfair offers, rather than simply shifting their acceptance thresholds. In early rounds, responders may evaluate unfair offers in a relatively undifferentiated way, focusing mainly on their own payoff. As they gain experience, however, they may become better able to infer whether the offer reflects intentional behavior or not. In particular, unfair offers from human proposers may be increasingly interpreted as intentional violations of fairness norms, which strengthens the willingness to reject them. By contrast, unfair offers generated by AI agents may continue to be perceived as less intentional or less blameworthy, making them relatively more acceptable even in later rounds. 

In T3, this effect may be further amplified because rejecting an unfair AI offer not only responds to the allocation itself but may also impose a cost on another human participant. As a result, learning may reinforce a divergence in how responders evaluate human and AI-generated offers over time, thereby sharpening the responder-side asymmetry observed in the later rounds.

While learning alone does not explain the origin of the asymmetry, the results here suggest that it shapes how the asymmetry emerges over time. In particular, learning appears to sharpen the responder-side asymmetry in later rounds by increasing the role of perceived intentionality and payoff consequences in the evaluation of unfair offers.

\subsection{Prior Beliefs}

\noindent In Survey A, which was conducted prior to the main task, participants were asked to predict the behavior of human participants and, in T2 and T3, the behavior of the AI agent (see Online Appendix C.1). These elicited beliefs allow us to compare prior expectations and expectation biases across treatments, and to explore whether they help explain the asymmetry in the main findings. 

Detailed descriptions and analyses are reported in Online Appendix E. Here, we briefly summarize the main findings.

\medskip
\noindent \textit{Predictions of Opening Offers.} There is no significant difference in predicted opening offers between T1 and T2, or between T2 and T3.

\medskip
\noindent \textit{Predictions of Agreement Timing.} Participants in T2 expected agreements to be reached slightly later than those in T1.

\medskip
\noindent \textit{Expectation Biases.} We measure expectation bias as the difference between predicted and actual outcomes. The results show that participants exhibited systematic expectation errors, particularly in the AI-related treatments. In particular, they (1) overestimated the human proposer's opening offer when bargaining with an AI agent; (2) overestimated the AI agent's opening offer to a human responder; and (3) underestimated the speed at which agreements were reached, especially when bargaining with AI agents. In addition, participants who overestimated human proposers' opening offers tended to make higher opening offers themselves, but were less willing to accept unfair offers as responders.

\medskip
\noindent \textit{Possible implications for the asymmetry.} Although these prior beliefs help illuminate some behavioral patterns on both the proposer side and the responder side, they do not appear to be the main reason why treatment effects differ between the two roles. In particular, controlling for prior beliefs does not overturn the main treatment patterns. Thus, prior beliefs are unlikely to be the key source of the asymmetry in the main results.

\subsection{Posterior Beliefs}

\noindent In Survey B, which was conducted after the main task, participants were asked to report their feelings during the task, their self-reported strategies, their perceptions of the AI agent, and, in T3, their understanding of the human-beneficiary payment rule (see Online Appendix C.2).

Detailed descriptions and analyses are reported in Online Appendix F. Here, we briefly summarize the main findings.

\medskip
\noindent \textit{Feelings.} Participants bargaining with an AI agent reported significantly lower anger, disappointment, and perceived respect than those bargaining with a human opponent. We find no significant evidence that the human-beneficiary manipulation changed these reported feelings relative to the pure human--AI treatment.

\medskip
\noindent \textit{Self-reported strategies.} Compared with human--AI bargaining, human--human bargaining was associated with a significantly higher tendency to report rejecting or retaliating against unfair offers. This pattern is consistent with the view that unfair offers elicited stronger punitive responses in human--human bargaining.

\medskip
\noindent \textit{Perceptions of bargaining with AI.} None of the AI-perception measures differs significantly between T2 and T3. 

\medskip
\noindent \textit{Understanding of the human-beneficiary payment rule.} The \textit{human-beneficiary payment rule} was generally well understood. However, although most participants understood the rule, many reported that it did not strongly affect their decisions. This suggests that the manipulation was psychologically meaningful, but only moderately salient.

\medskip
\noindent \textit{Possible implications for the asymmetry.} These posterior responses suggest that the asymmetry is unlikely to be driven by a broad change in how participants perceived the AI agent itself. Instead, they are more consistent with a decision-specific interpretation: participants reported weaker emotional reactions in human--AI bargaining, while the human-beneficiary manipulation may have changed the perceived payoff consequences of rejecting unfair AI offers without substantially changing participants' general perceptions of the AI agent.

\section{Conclusion}
\label{section6}
\noindent As LLM-based AI negotiation systems increasingly appear in real-world commercial settings, understanding how humans bargain with AI counterparts has become an important question. In this paper, we conducted a laboratory experiment to compare human–human bargaining and human–AI bargaining in a 3-stage alternating-offer game. In addition, motivated by the fact that many real-world AI bargaining systems act on behalf of firms or other humans, we introduced a human-beneficiary manipulation in which the AI agent’s earnings could affect another participant’s final payoff. Our aim was to examine, first, whether social-preference considerations become weaker in human–AI bargaining than in human–human bargaining, and second, whether linking the AI agent’s payoff to another human participant can partially restore such considerations.

Our results show that, in an alternating-offer bargaining game with at most three stages, the speed of reaching agreement does not differ significantly between human–human bargaining and human–AI bargaining. However, when the AI agent’s earnings are linked to another human participant, agreements are reached significantly earlier. This suggests that the human-beneficiary manipulation promotes earlier agreement in human–AI bargaining and is therefore consistent with a partial restoration of social-preference considerations at the aggregate level.

More importantly, we find a clear asymmetry between proposer behavior and responder behavior. On the proposer side, human participants offer more to human opponents than to AI agents, but the human-beneficiary manipulation does not significantly increase offers to AI agents. On the responder side, human participants do not differ significantly in their willingness to accept unfair opening offers from human versus AI proposers. However, when the AI agent’s earnings are linked to another human participant’s payment, responders become significantly more willing to accept such unfair offers. These findings indicate that treatment effects in human–AI bargaining are role-dependent. The weakening of social-preference considerations in human–AI bargaining appears more clearly on the proposer side, whereas the restoring effect of the human-beneficiary manipulation appears more clearly on the responder side.

We discuss several possible explanations for this asymmetry. Our additional analyses suggest that learning, prior beliefs, and posterior beliefs do not provide the main explanation for the difference between proposer-side and responder-side treatment effects. Instead, our analysis of FMA suggests one possible interpretation. In our bargaining environment, proposers enjoy a strong structural advantage, especially in human–AI bargaining. As a result, the human-beneficiary manipulation may not be strong enough to generate a significant increase in proposers’ offers. By contrast, responders do not benefit from the same FMA, and their accept-or-reject decisions make the social consequences of unfair offers more directly relevant. This may explain why the human-beneficiary manipulation appears more clearly on the responder side.

Overall, our findings contribute to the growing literature on human–AI bargaining, especially bargaining with LLM-based AI agents, which is becoming increasingly realistic in practice. More broadly, the results suggest that human responses to AI bargaining systems cannot be characterized by a single uniform shift in behavior. Instead, such responses depend on the strategic role of the human participant and on whether the AI agent’s payoff is socially consequential. These findings may also help inform the design of AI negotiation systems by highlighting the importance of users’ social and psychological reactions to AI counterparts, thereby helping firms better anticipate human responses and design negotiation agents that more effectively pursue organizational objectives.

This study has several limitations. First, although our game allows for up to three stages, the limited number of observations beyond Stage 1 means that most of our analysis focuses on opening offers and responses to opening offers, while evidence on later-stage counteroffers remains limited. Second, we examine only one combination of discount factors, corresponding to one of the treatments in \citet{OchsRoth1989}. Because this parameterization strengthens the first-mover advantage of the proposer, future research should investigate whether alternative discount-rate combinations lead to different patterns of asymmetry. Third, the AI agent used in our experiment was prompted only with the rules of the game and was not strategically tuned through prompt engineering. In real-world applications, however, AI negotiation systems are often carefully designed and optimized for particular goals. Future research could therefore examine how different prompting strategies or AI behavioral styles affect human responses in bargaining environments.

\newpage
\bibliography{Ref/Ref2604,Ref/refer_pilot}
\newpage

\end{document}